# Different factors determining Motor Execution and Motor Imagery performance in a serial reaction time task with intrinsic variability

**Patricia Silva de Camargo[1]\*, Paulo Roberto Cabral-Passos[2], André Frazão Helene[3].**

[1]Institute of Psychology, University of São Paulo (IP-USP), São Paulo, Brazil.

[2]Faculty of Philosophy, Science and Letters Ribeirão Preto, Department of Physics and Mathematics, University of São Paulo (FFCLRP-USP), Ribeirão Preto, Brazil.

[3] Institute of Biosciences, Department of Physiology, University of São Paulo (IB-USP), São Paulo, Brazil.

**\*Correspondence:**
Corresponding Author: Patricia Silva de Camargo
E-mail: patysdcamargo@gmail.com

**Dedication:** to Antonio Galves, CEPID NeuroMat's scientific director.

**Keywords:** motor imagery; imagery; learning; probabilistic sequence; serial reaction time task.

**Abstract:** Motor imagery corresponds to the mental practice of simulating visual and kinesthetic aspects of a given motor task. This practice shares a similar neural substrate and correlated temporal scale with motor execution. Besides that, it can lead to performance improvements in the actual execution of the imagined task. Therefore it is important to understand functional differences and equivalences between motor imagery and motor execution. To tackle that we employed a finger-tapping serial reaction time task in two groups of participants, a Motor Execution (n=10) and a Motor imagery (n=10). The sequence of stimuli defining the task had 750 items composed of three distinct auditory stimuli. Also, this sequence had some intrinsic variability making some of the next items unpredictable. Each auditory stimulus was mapped to a single right hand finger in the Motor Imagery group. The Motor imagery group indicated the end of the imagination with a single response using the left hand. The results show improvement in performance of the Motor Imagery group throughout the task and that the duration of the motor imagery, indirectly measured by reaction times, are influenced by distinct factors than those of Motor Execution.

**Introduction**

Motor imagery corresponds to the mental practice of simulating visual and kinesthetic aspects of a given motor task (Helene and Xavier, 2006; Schuster et al., 2011). Evidence shows that sensorimotor areas such as the primary motor cortex, cerebellum, and basal ganglia present an activation during motor imagery that resembles that of motor execution (Jeannerod, 2001; Dickstein and Deutsch, 2007; Munzert, Lorey, and Zentgraf, 2009; Henschke and Pakan, 2023). Besides that, the duration of motor imagery and motor execution is also correlated (Papaxanthis et al., 2002). Therefore, it is reasonable to assume that motor imagery and execution share similar mechanisms and might lead to similar outcomes, such as learning (Helene and Xavier, 2006; Schuster et al., 2011). This assumption is supported by improvements in performance shown in motor imagery studies (Driskell et al., 1994; Collet et. al., 2011, Barclay et. al., 2020; Liu et. al., 2024).

The essence of mental practice is to induce an experience without the original input of external stimuli (Tian and Poeppel, 2010). Motor *emulation* theory (MET) proposes that the brain presents emulators and plan encoders. The purpose of the former would be to provide predictions to the latter, enabling corrections and adjustments (Hurst and Boe, 2022). According to the MET, these emulators still provide forward models of sensory predictions during Motor Imagery (Grush, 2004; Wolpert, Ghahramani, and Jordan, 1995; Hurst and Boe, 2022). In both, motor execution and motor imagery, the predictions from the emulators are based on somatosensory consequences estimated from the efferent copy of motor commands (Desmurget and Sirigu, 2009). Efferent copies contain information about the motor execution, sensory consequences, and location of the action (afferent proprioceptive input), and they are directly involved in error processing, motor control, motor execution, visual and auditory perception, posture, and tactile perception (Stock et al., 2013).

To tackle learning associated with motor imagery, in the present article, we use a paradigm introduced by Duarte et al. (2019) to study how humans learn regularities in sequences of events with unpredictable steps. Humans must often deal with sequences that are unpredictable at some level and have to make decisions based on them, therefore this paradigm has a strong ecological appeal. In this paradigm, stimuli are generated based on a probabilistic context tree model, henceforth referred to just as a context tree. Basically, a context tree is a set of contexts and associated probability measures. In a sequence following a context tree, each new element is generated based on a suffix of the past sequence, which may vary in length (Rissanen, 1983). In turn, each context of the sequence presents an associated set of probability measures. If the sequence is composed of an alphabet, say A, B, and C, these probabilities dictate how often A, B, and C come after the context. A sequence can be infinitely long but with a finite number of contexts of finite length and for each context, the probabilities are fixed. So learning the sequence translates to inferring the contexts and probability measures of the sequence (Hernández et. al., 2021).

The current study introduced a serial reaction time task in which participants should perform a finger tapping according to a sequence of stimuli generated by a context tree. The task was performed by two groups, Motor Execution and Motor Imagery. We hypothesized that (1) participants performing the motor imagery would improve performance after training and (2) the duration of the motor imagery would be influenced by factors that differ from the ones that determine the reaction times in motor execution.

**Materials and Methods**

*1. Participants*

The study included two groups of healthy right-handed participants capable of performing imaginative tasks: the motor execution group with 10 participants (6 women, mean age/standard deviation of 27.1±4,18 years) and the motor imagery group with 10 participants (5 women, mean age/standard deviation of 26.4±4,93 years). Refer to Table 1 for detailed characteristic information about all the volunteers. The experiment occurred at the Cognition Science Laboratory in the Institute of Biosciences (IB-SP) of the University of São Paulo. All participants were either undergraduate or graduate students at the University of São Paulo. The study received approval from the local ethics committee (CAAE 15274718.8.0000.5464, Institute of Biosciences at the University of São Paulo - IB-USP), and all participants signed the informed consent.

Participants were selected according to the following criteria. They should be over 18 years old, right-handed, and have the ability to perform imaginative tasks. They also should present no history of neurological dysfunctions, spinal fractures, orthopedic, vascular, or muscular dysfunctions affecting the upper limb. The handedness dominance was evaluated using the *Edinburgh Handedness Inventory* (Oldfield, 1971). Participants with scores greater than 50 were considered right-handed. The ability to perform imaginative tasks was evaluated using the *Kinesthetic and Visual Motor Imagery Questionnaire - KVIQP-10, Brazilian version* (Demanboro et al., 2018)

**Table 1: Demographic and questionnaire information of the participants separated into groups. Means and standard deviations are given.**

|  | Motor Execution | Motor Imagery | Total |
| --- | --- | --- | --- |
| Age (years) | 27.1±4.18 | 26.4±4.93 | 26.75±4.46 |
| Gender | F(6); M(4) | F(5); M(5) | F(11); M (9) |
| Edinburgh Inventory | 82.73±13.75 | 86.51±15.3 | 84.62±14.29 |
| KVIQP-10V | 35.1±4.79 | 26.1±8.91 | 30.6±8.35 |
| KVIQP-10K | 32.9±6.45 | 25.8±6.98 | 29.35±7.49 |

F = female; M = male; KVIQP-10V = Visual Motor Imagery Questionnaire (Brazilian version); KVIQP-10K = Kinesthetic Motor Imagery Questionnaire (Brazilian version).

## 2. Experimental task

The experiment used *Psychtoolbox-3 (Psychophysics Toolbox Version 3 - PTB-3)*, software version 3.0.17 Beta to present auditory stimuli. The auditory stimuli set were labeled as one, two, and three indicating which action the participant should take. They were recorded using a standard microphone, and the volume was adjusted individually for each participant to ensure comfort.

The experiment was a serial reaction time task. The sequence of stimuli was determined by a context tree whose alphabet was 1 (indicator), 2 (middle), and 3 (ring). This context tree was previously tested in another study (Uscapi, 2020). A sample of the sequence generated by this context tree is shown in Figure 1. For a friendly explanation of how to generate sequences using context trees, we refer the reader to Cabral-Passos et al. (2024). Responses were recorded using a standard keyboard, with the left arrow corresponding to 1, the down arrow to 2, and the right arrow to 3. The events could be either fixed stimuli (with a 100% chance of occurrence) or variable stimuli (with a 26% chance of occurrence for the event 2 after another 2 and a 74% for the event 3).

The events could be either deterministic or non-deterministic. The deterministic events are labeled with an F (of fixed) in Figure 1b, they correspond to 1 that follows the 3 (F1) and the 2 that follows the 1 (F2). On the other hand, non-deterministic events are labeled with a V (of variable) in Figure 1b, they correspond to the 2 that follows a 2 (V2) and the 3 that follows a 2 (V3). The non-deterministic events V2 and V3 have 26% and 74% percent chance of occurrence, respectively.

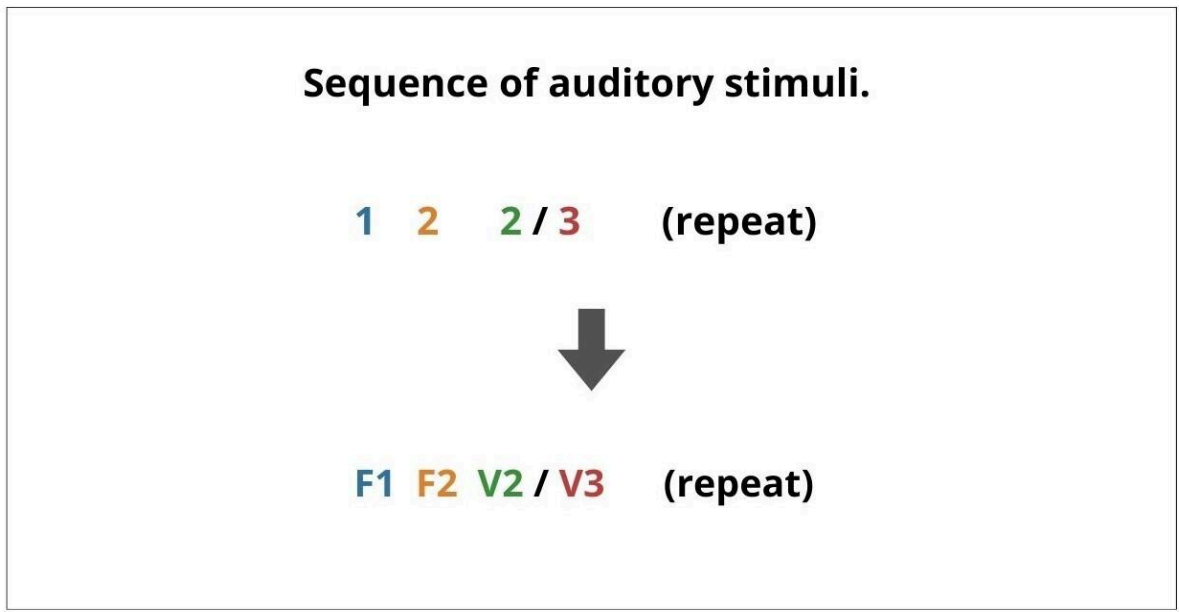

**Figure 1: Graphical representation of the sequence of stimuli in the task.**
The sequence of auditory stimuli where "1" and "2" with 100% probability of occurrence are fixed events in the sequence (called F1, blue, and F2, orange, posteriorly); the occurrence of "2" (26% probability) or "3" (74% probability) are variable events (V2, green, and V3, red, respectively).

### 3. Experimental Protocol

Participants performed the task seated in front of a 17-inch LCD monitor that was positioned 60cm away and aligned to the participant's eyes. The room in which the task was performed remained silent throughout the task. The researcher provided the instructions and clarified any questions to the participants before the task initiation.

#### a. Familiarization Phase

Participants were presented with an auditory stimulus of a number (1, 2, or 3) during each trial. Each number corresponds to a specific finger of the right hand, with the index finger associated with the number 1, the middle finger with the number 2, and the ring finger with the number 3. Participants were instructed to press the corresponding key on the keyboard as quickly as possible upon identifying the number. Participants were informed that if they pressed the key before the auditory stimulus a beep would sound indicating that the current trial was invalid before moving to the next trial.

The reaction time for each trial was defined as the time from the end of the auditory cue until the beginning of the motor response (see Figure 2). This phase consisted of only six trials, with two trials associated with each of the fingers and randomly assigned order.

#### b. Test Phase

During the test phase, participants were presented with 750 auditory stimuli divided into 5 blocks of 150 trials each. Table 2 the frequency associated with F1, F2, V2, and V3 for each block.

**Table 2: Quantity of each event in the sequence of auditory stimuli throughout the experimental blocks.**

|    | B1 | B2 | B3 | B4 | B5 | Total |
|----|----|----|----|----|----|-------|
| F1 | 50 | 50 | 50 | 50 | 50 | 250   |
| F2 | 50 | 50 | 50 | 50 | 50 | 250   |
| V2 | 16 | 7  | 13 | 14 | 15 | 65    |
| V3 | 34 | 43 | 37 | 36 | 35 | 185   |

B1 = first experimental block; B2 = second experimental block; B3 = third experimental block; B4 = fourth experimental block; B5 = fifth experimental block.

Participants were instructed to keep their eyes closed until the end of each block indicated by the sound of two consecutive "beeps". In the motor execution group, participants were required to press the correct key with the corresponding finger of the right hand immediately after identifying the number (1, 2, or 3). There was no register of any key presses during the stimulus presentation. If a participant missed the key press or pressed the wrong key, the program waited for the correct response before proceeding to the next trial (mean and standard deviation of missed key presses: $0.001 \pm 0.033$). In the motor imagery group, participants were instructed to visually imagine themselves performing the finger-tapping task in first-person perspective. Once the motor imagery was completed, they were instructed to press the "spacebar key" with the left index finger as illustrated in Figure 2.

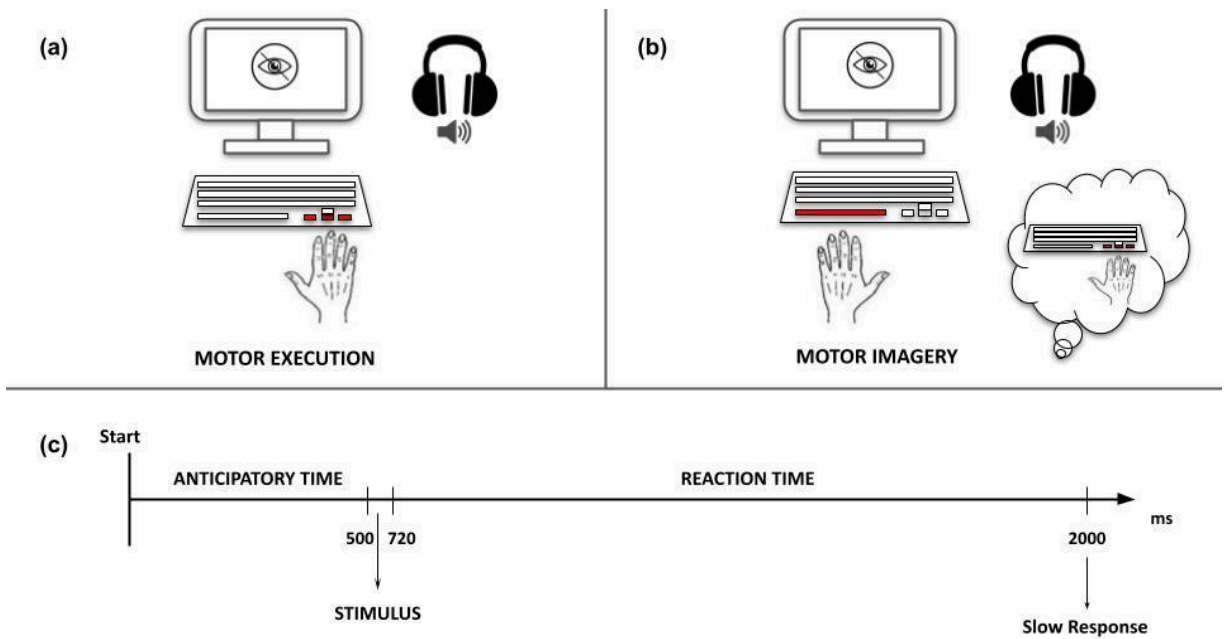

**Figure 2: Schematic representation of each trial under the conditions of Motor Execution and Motor Imagery.**
Schematic representation of each trial under the conditions of (a) Motor Execution and (b) Motor Imagery. (c) Trial schematic arrangement containing the anticipatory time (500ms before the stimulus), the average duration of the auditory stimulus (220ms), the reaction time that depends on the individual's motor response, and the duration of what was considered a slow response (2000ms).

### 4. *Statistical Analysis*

All statistical analyses were performed using the Python programming language. Additionally, Statsmodels and Pingouin Python packages were used for statistical tests (Seabold and Perktold, 2010; Vallat, 2018). Two-way ANOVA designs were used to test differences in distributions of mean reaction times across blocks, events, and groups. Then, a post-hoc pairwise comparison tests were performed to identify which conditions differed. The mean reaction times during the test phase were compared across groups in a 2-way mixed ANOVA. "Group" was considered as a between-subjects factor, while "Block" and "Event" were regarded as within-subjects factors. Only correct reaction times smaller than 2 seconds were included in the analysis. All code, output and description is available in the following link:https://github.com/PauloCabral-hub/third_party_work/blob/main/data_analysis_routine.ipynb

## Results

### 1. *Performance Analyses Associated to the Probabilities of the Events*

Mean reaction times were used as indirect measures of performance in all of the following analyses. Figure 3 shows the distribution of the mean reaction times for F1, F2, V2, and V3 events in each of the five blocks of the Motor Execution (left) and Motor Imagery (right) groups. Therefore, each distribution on the graph has a sample size of 10. For both groups, a 2-way repeated measures ANOVA was used to verify significant differences across blocks and events. Before the 2-way repeated measures ANOVA, the distributions of each event and block were tested for normality and sphericity adopting a significance level of 0.01. The Shapiro-Wilk test indicated a violation of normality in block 1 (Shapiro and Wilk, 1965). Given that, the values of the distributions were rank-transformed and tested for normality again. After transformation, no violation was detected. The *p*-values of the 2-way repeated measures ANOVA were corrected using Greenhouse–Geisser correction in case of violations of sphericity (Greenhouse and Geisser, 1959).

For Motor Execution group, the 2-way repeated measures ANOVA revealed significant effects for the factor Event ($F_{3,27}$=5.569, $p<0.012$), but not for Block ($F_{4,36}$=1.252, $p$=0.310) and no interaction between Block and Event ($F_{12,108}$=1.095, $p$=0.372). Post-hoc pairwise comparison tests showed differences for the Events F1 and F2 ($p<0.01$) (see supplementary table S1 for details).

For the Motor Imagery group, the ANOVA revealed significant effects of Block ($F_{4,36}$=17.456, $p<0.001$) and Event ($F_{3,27}$=6.095, $p<0.016$), but no interaction between Block and Event ($F_{12,108}$=1.734, $p$=0.189). The list of significant post-hoc pairwise comparisons with the respective Bonferroni corrected *p*-values are available in the supplementary table S2 (Bonferroni, 1936). Significant pairwise comparisons for specific distributions are depicted in Figure 3 as bars of different thicknesses, thin bars for $p < 0.05$ and thick bars for $p < 0.01$. The mean reaction times for V2 and V3 decreased across the blocks. The mean reaction times in V2 was higher during blocks, potentially due to the smaller probability of this event.

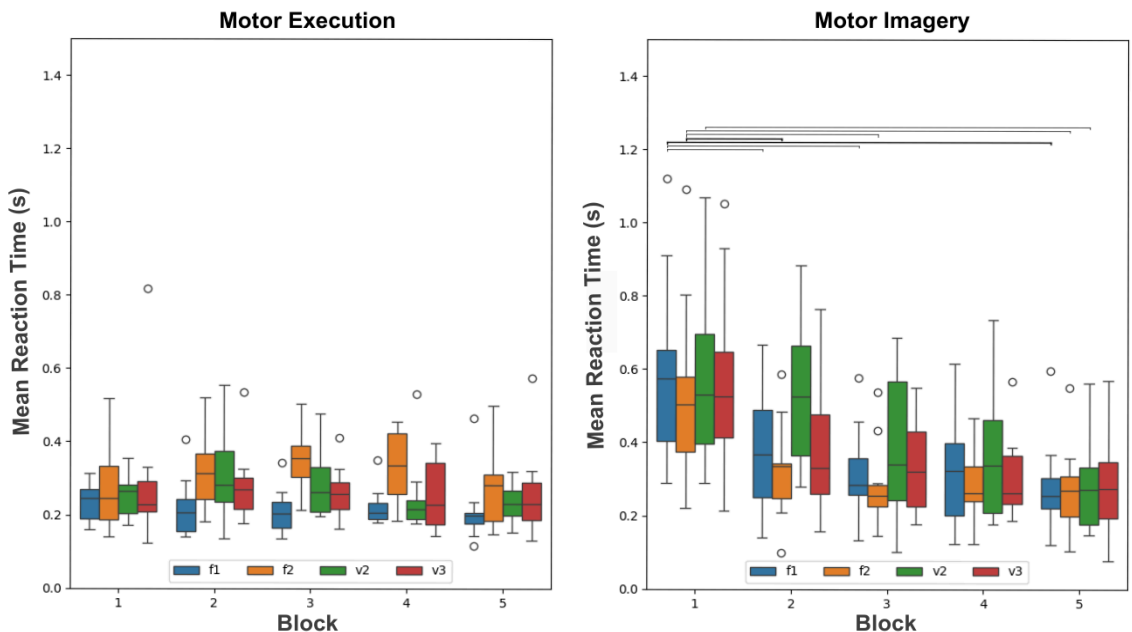

**Figure 3: Performance analysis (mean reaction times) of the Motor Execution and Motor Imagery as a function of Block and Event.** Distributions of the mean reaction times for the events F1 (blue), F2 (orange), V2 (green), and V3 (red) across blocks for the Motor Execution and Motor Imagery groups. The line between the first and third quartiles indicate the median value. Bars on top of the distributions indicate the statistical significance of the post-hoc tests (thin bars for $p < 0.05$ and thick bars for $p < 0.01$). Outliers are indicated as small circles.

A 2–way mixed ANOVA was used to compare both Groups across Events. Figure 4 shows the distributions of mean reaction times of Motor Execution (left) and Motor Imagery (right) disregarding block information. Before applying the 2–way mixed ANOVA, the distributions were tested for normality, homogeneity of variance, and sphericity. Given the violation in normality indicated by the Shapiro-Wilk test, the values were rank transformed and tested again for the three assumptions. The Levene test was used to test the homogeneity of variance (Levene, 1960). The 2–way mixed ANOVA indicated significant differences for the between-subjects factor Group ($F_{1,18}=7.353$, $p < 0.05$), but not for the within-subjects factor Event ($F_{3,54}=3.849$, $p > 0.05$). The test also indicated significant interaction of Group and Event ($F_{3,54}=7.172$, $p < 0.01$). The list of significant post-hoc pairwise comparisons with the respective Bonferroni corrected *p*-values are available in the supplementary table S3. Significant pairwise comparisons for specific distributions are depicted in Figure 4 as bars of different thicknesses, thin bars for $p < 0.05$ and thick bars for $p < 0.01$. It is evident from the figure that except for F2, all events differed between groups. On the other hand, only some events differ within each group.

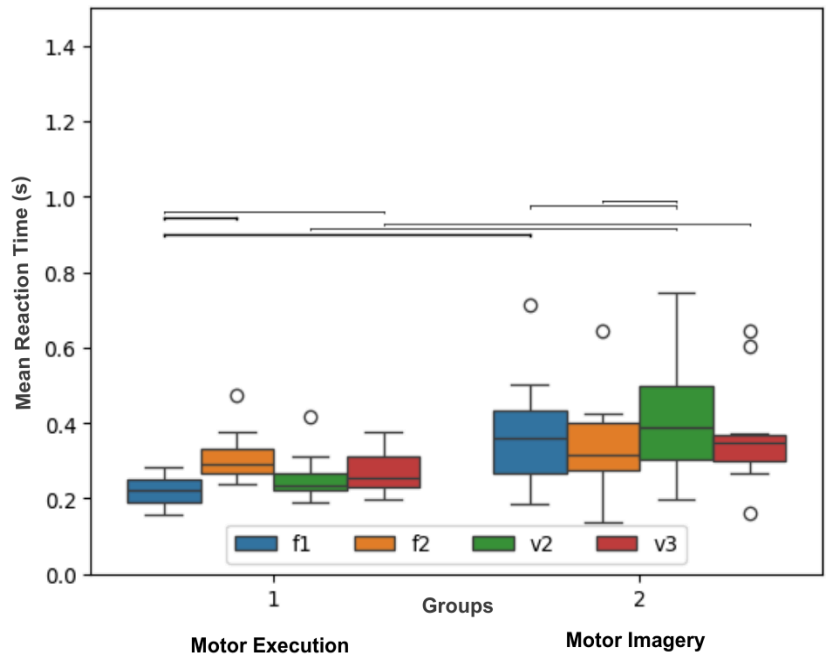

**Figure 4: Performance of the Motor Execution and Motor Imagery groups as a function of Event.**
Distribution of the mean reaction times for events F1 (blue), F2 (orange), V2 (green), and V3 (red) across groups (Motor Execution - Left, Motor Imagery - Right). The line between the first and third quartiles indicate the median value. Bars on top of the distributions indicate the statistical significance of the post-hoc tests (thin bars for $p < 0.05$ and thick bars for $p < 0.01$). Outliers are indicated as small circles.

### 2. *Performance Analysis as a Function of the Impact of Last Variable Event*

Figure 5 shows the distribution of mean reaction times for each Event given the immediate Last Variable Event for both groups, Motor Execution (left) and Motor Imagery (right). That is, for both groups, the four distributions on the left indicate the mean reaction times following a V2 for F1, F2, V2 (itself) and V3. On the other hand, the four distributions on the right indicate the mean reaction times following a V3 for F1, F2, V2, and V3 (itself). To test for differences of the Events across the Last Variable Event a 2–way repeated measures ANOVA was applied to the distributions of the mean reaction times of each group. There were no indications of violation of normality by the Shapiro-Wilk test. Violations of sphericity were treated with Greenhouse-Geisser correction. For the Motor Execution group, no difference was observed for the factor Event ($F_{3,27}$=3.313, $p > 0.05$) and Last Variable Event factor ($F_{1,9}$=1.009, $p > 0.05$). On the other hand, a significant interaction of both factors was observed ($F_{3,27}$=6.169, $p < 0.01$). The list of significant post-hoc pairwise comparisons with the respective Bonferroni corrected *p*-values are available in the supplementary table S4. Significant pairwise comparisons for specific distributions are depicted in Figure 5 as bars of different thicknesses, thin bars for $p < 0.05$ and thick bars for $p < 0.01$. As depicted on Figure 5 (left), the mean reaction times for F1 are smaller after the last variable event V2 than after V3. On the other hand, the converse occurs for F2. This contrast suggests that the anticipation of the variable event in F2 may increase the reaction times.

Similarly, for the Motor Imagery group, no difference was observed for the factor Event ($F_{3,27}$=3.334, $p > 0.05$) and Last Variable Event factor ($F_{1,9}$=1.526, $p > 0.05$). However, a significant interaction of both factors was observed ($F_{3,27}$=13.216, $p < 0.01$). The list of significant post-hoc pairwise comparisons with the respective Bonferroni corrected *p*-values are available in the supplementary table S4. Significant pairwise comparisons for specific distributions are depicted in Figure 4 as bars of different thicknesses, thin bars for $p < 0.05$ and thick bars for $p < 0.01$. The list of significant post-hoc pairwise comparisons with the respective Bonferroni corrected *p*-values are available in the supplementary table S5. Significant pairwise comparisons for specific distributions are depicted in Figure 5 as bars of different thicknesses, thin bars for $p < 0.05$ and thick bars for $p < 0.01$. As depicted on Figure 5 (right), the mean reaction times for V2 are smaller after the last variable event V2 than after V3. On the other hand, the converse occurs for V3, that is, mean reaction times are larger after V2. This contrast does not appear in the motor execution group. Besides that, no differences are verified between F1 and F2 following V2 and V3.

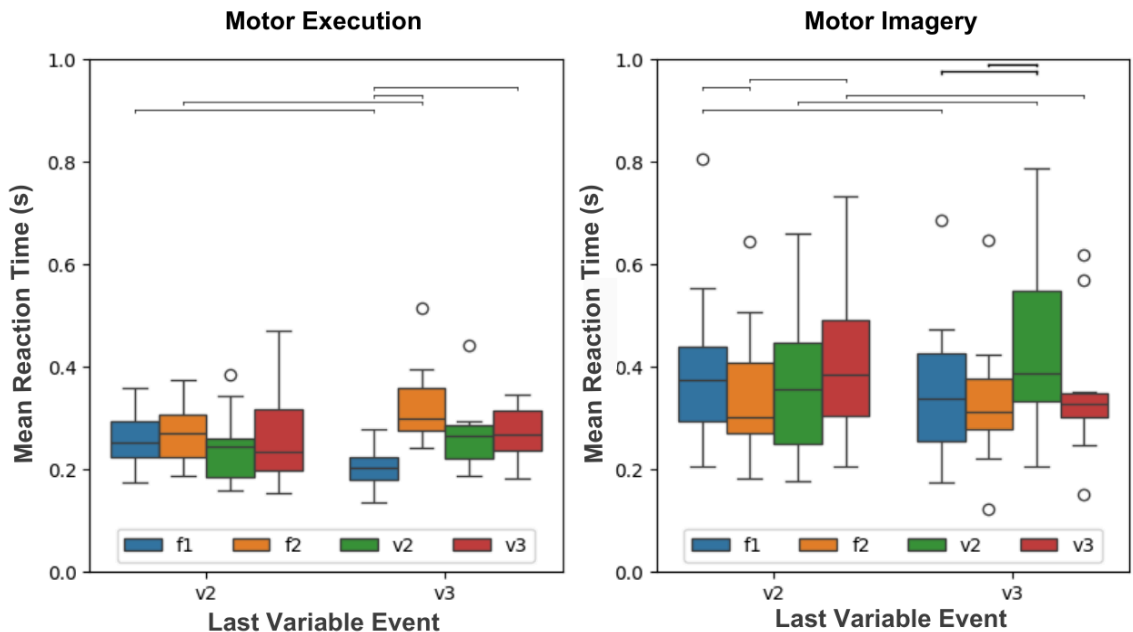

**Figure 5: Performance of the Motor Execution and Motor Imagery groups for each Event as a function Last Variable Event.** Distributions of mean reaction times for the events F1 (blue), F2 (orange), V2 (green), and V3 (red) after the last variable event (V2 and V3) for the Motor Execution and Motor Imagery groups. The line between the first and third quartiles indicate the median value. Bars on top of the distributions indicate the statistical significance of the post-hoc tests (thin bars for $p < 0.05$ and thick bars for $p < 0.01$). Outliers are indicated as small circles.

**Discussion**

The current study was developed to investigate functional equivalences/differences between motor imagery and motor execution. The study provides evidence of how sequence structures with intrinsic variability are processed by the nervous system during motor imagery. Given that motor imagery enhances learning of motor sequences and benefits the development of fine motor skills (Sobierajewicz et al., 2016), we aimed at investigating how motor execution and motor imagery differs during a structural learning task. We used reaction time as an indirect measure of performance. The finger-tapping serial reaction time task used followed a structure determined by the same context tree across all experimental blocks and presented events with different probabilities resulting in a sequence with intrinsic variability. The results showed that (1) performance improves in the Motor Imagery group in the given task and (2) the duration of motor imagery indirectly measured by reaction times are influenced by distinct factors than those of Motor Execution.

The current study showed that the mean reaction times of the Motor Imagery group decreased across the task blocks (Figure 3, right). Surprisingly, there was no significant reduction in reaction times across blocks for the Motor Execution group (Figure 4, left). These differences can be attributed to the lack of the somatosensory feedback associated with the three-finger-mapped executions in the Motor Imagery group. The subjects in the Motor Imagery group were instructed only to imagine performing three-finger-mapped execution, not actually performing it. In the Motor Execution group, the absence of significant reduction of the mean reaction times across blocks may be related to the simplicity of the task. Unfortunately, there is no consensus on how to measure the complexity of a sequence (see Hernandez et al., 2024), but a similar sequence was used in Cabral-Passos et al. (2024) with a high percentage of correct predictions (see paper Figure 4). If that is the case, learning occurs in a small time interval in the Motor Execution group as compared to the Motor Imagery group and the lower limit of reaction times is reached sooner in the task.

The 2-way repeated measures ANOVA applied to the data in Figure 3 indicated that both groups were affected by the probabilistic structure of the sequence (see paragraphs 2 and 3 in results section). This is even more evident from the results in Figure 4. Also, despite the correlation between the duration of motor execution and motor imagery presented in Papaxanthis et al. (2002), the 2-way mixed ANOVA applied to the data in Figure 4 indicated significant interaction of the factors Group and Event. This indicates that the mean reaction times have a different behaviour in each group. Interestingly, the mean reaction times of F1 and F2 differ in the Motor Execution group, but not in the Motor Imagery group. One possible explanation for this finding is that the expectation regarding the next event, V2 (26%) or V3 (74%), is more pronounced in the Motor execution group given the associated, more stimulus specific, motor preparation.

Another important finding of the current study is related to the data in Figure 5. It was shown a significant interaction of the factors Event and Last Variable Event in both groups. This result indicates that the behaviour of the events also changes as a function of the last variable event in serial reaction tasks. In the Motor Imagery group, it is possible to see that the mean reaction times are smaller for V2 following V2 as compared to V2 following V3. Similar, for the same group, mean reaction times are smaller for V3 following V3 as compared to V3 following V2. This suggests that motor planning is continuously taking into account the last variable event even in Motor Imagery.

Methods used to monitor brain activity have produced evidence that similar brain areas are activated during Motor imagery and Motor execution (Jeannerod and Decety, 1995; Crammond, 1997). Other structures such as the descending motor pathways (Pascual-Leone et al., 1995), the autonomic nervous systems (Decety et al., 1993) and even spinal reflex pathways (Oishi et al., 1994) seem to be affected by the practice. Furthermore, improvements of motor function induced by motor imagery have been found in healthy subjects (Schuster et al., 2011; Park et al., 2014; Ridderinkhof and Brass, 2015; Di Rienzo et al., 2016; Romano-Smith et al., 2018; Lotze, 2013) and subjects in motor rehabilitation (Mulder, 2007). Therefore, it makes sense to understand how learning occurs in motor imagery as a way to develop and test motor control and motor emulation theories (Wolpert and Ghahramani, 2000; Kilteni et al., 2018; Grush, 2004; Hurst and Boe, 2022) that can help produce guidelines for participants to get the most out of the practice.

**Conclusion**

In summary, our results indicate that both Motor Execution and Motor Imagery groups were affected by the intrinsically variable structure of the sequence of stimuli. The Motor Imagery group showed a marked improvement across blocks for the different events. Moreover, the mean reaction times showed different behaviour in both groups in respect to event, block and last variable event. Further studies are necessary to explore which parameters determine these differences and the specific neural activity patterns and neural substrates that differentiate motor execution and motor imagery.

**Supplementary material**

Supplementary material available at:
https://github.com/PauloCabral-hub/third_party_work/blob/main/Camargoetal2024_sup.pdf

**Conflict of Interest:**

The authors declare that the research was conducted in the absence of any commercial or financial relationships that could be construed as a potential conflict of interest.

**Author Contributions:**

All authors contributed equally to this manuscript.

**Funding:**

The following Brazilian research agencies supported this study: "Coordenação de Aperfeiçoamento de Pessoal de Nível Superior - Brasil - CAPES" (Finance Code 001) and "Research, Innovation and Dissemination Center for Neuromathematics - NeuroMat" (FAPESP - process number 2013/07699-0).

**Acknowledgments:**

We thank the Brazilian research agencies, especially the CEPID NeuroMat, for the funding support. We also thank Antonio Galves (*in memoriam*), NeuroMat's scientific director, for all the scientific support and leadership during this research.